\documentclass[pra,superscriptaddress,twocolumn]{revtex4}

\usepackage[SquareTraceBrackets]{quantum}
\usepackage{graphicx,bm,natbib,upgreek,amssymb,amsmath}

\usepackage[colorlinks=true,citecolor=black,linkcolor=black,urlcolor=black]{hyperref}

\usepackage[T1]{fontenc}
\usepackage[scaled]{helvet}

\begin{document}

\title{\large\textsf{\textbf{The effect of component variations on the gate fidelity in linear optical networks.}}}

\author{Jonathan Crickmore}
\affiliation{Department of Physics \& Astronomy, University of Sheffield, Sheffield S3 7RH, United Kingdom}
\affiliation{Department of Physics, Heriot-Watt University, Edinburgh, United Kingdom}

\author{Jonathan Frazer}
\affiliation{Department of Physics \& Astronomy, University of Sheffield, Sheffield S3 7RH, United Kingdom}

\author{Scott Shaw}
\affiliation{Department of Physics \& Astronomy, University of Sheffield, Sheffield S3 7RH, United Kingdom}

\author{Pieter Kok}\email{p.kok@sheffield.ac.uk}
\affiliation{Department of Physics \& Astronomy, University of Sheffield, Sheffield S3 7RH, United Kingdom}

\begin{abstract}
 \noindent We investigate the effect of variations in beam splitter transmissions and path length differences in the nonlinear sign gate that is used for linear optical quantum computing. We identify two implementations of the gate, and show that the sensitivity to variations in their components differs significantly between them. Therefore, circuits that require a precision implementation may benefit from additional circuit analysis of component variations to identify the most practical implementation. We suggest possible routes to efficient circuit analysis in terms of quantum parameter estimation.
\end{abstract}

\date{\today}

\maketitle  

\section{Introduction}

\noindent
Optical networks are important for a wide variety of applications, from conventional optical routers and classical optical computing \cite{shlomi10} to quantum communication networks, linear optical quantum computing and optical metrology \cite{klm01,kok07,koklovett10}. The physical system that underpins all of these networks is the \emph{multi-mode interferometer}. It is a collection of passive optical elements such as beam splitters, phase shifters and polarisers, as well as active elements such as optical squeezers, photodetectors and switches. These networks can be implemented in bulk optics, fibre optics, or on chip. The typical applications such as optical quantum computing, quantum imaging and quantum metrology all require an extremely high precision in the optical elements. However, in any practical implementation there will be significant variations in the interferometer elements (sometimes exceeding 10\% of the specified value). Depending on the application, such variations may be critically detrimental to the operation of the interferometer. Furthermore, the variations in some elements will have a much greater effect on the functionality of the interferometer than those of others. It is therefore key to improve our understanding of element sensitivity of optical circuits.

\begin{figure}[b!]
\includegraphics[width=8.5cm]{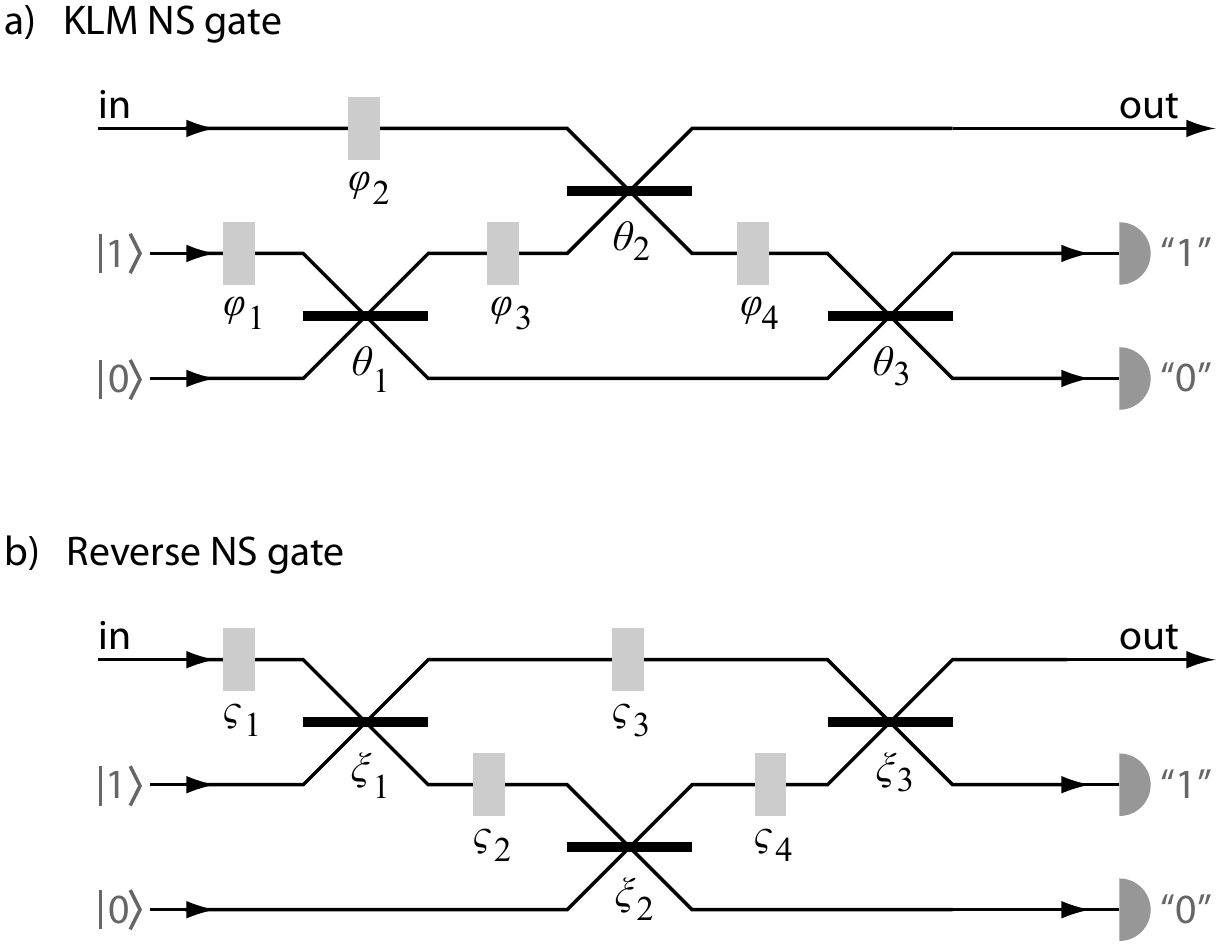}
\caption{two designs for the NS gate: a) the KLM NS gate, and b) the Reverse NS gate. The two circuits have the same complexity in terms of components, input states and detectors, and they have the same probability of successfully applying an NS gate ($\smash{p_{\rm success} = \frac14}$). However, the two circuits differ dramatically in the way they respond to variations in their component characteristics.}
\label{fig:ns}
\end{figure}

In this paper, we explore the sensitivity of the nonlinear sign (NS) gate used in linear optical quantum computing \cite{klm01} as an example of circuit variation analysis. We consider two versions of this gate with identical circuit complexity in terms of the number of optical elements, input states and detection devices, and operating at the same success probability (see Fig.~\ref{fig:ns}). We will find that the design of the gate has significant implications for the process fidelity's sensitivity on variations in the components. This means that any optical interferometer design will have to be tested against alternative designs for the best performance given realistic components. In Section~\ref{sec:ns} we present two different version of the NS gate and in Section \ref{sec:components} we study the effect of components variations. Section \ref{sec:multi} looks at the broader connection to multi-parameter estimation theory, and we conclude with a brief discussion in Section \ref{sec:disc}.

\section{Two NS gate designs}\label{sec:ns}

\noindent
The NS gate is a key component in the original proposal by Knill, Laflamme and Milburn (KLM) for a quantum computer constructed from linear optical elements, single photon sources and photodetection \cite{klm01}. The main operation of the gate is to induce a nonlinear phase shift $U_{\rm NS}$ on an optical mode defined by
\begin{align}\label{eq:ns}
 \alpha\ket{0} + \beta\ket{1} + \gamma\ket{2} \underset{U_{\rm NS}}{\longrightarrow} \alpha\ket{0} + \beta\ket{1} - \gamma\ket{2}\, ,
\end{align}
where $\ket{n}$ is the state of $n$ photons in the input mode and $(\alpha,\beta,\gamma)$ are complex amplitudes normalised to 1. The action of the NS gate on photon number states $\ket{n}$ with $n>2$ is not defined, and allows for additional freedom in the construction of $U_{\rm NS}$. 

The optical circuit for the original NS gate is shown in Fig.~\ref{fig:ns}a. It is a simple circuit that lends itself well to analysis. However, the circuit is not unique. We can define a ``reverse'' NS gate, shown in Fig.~\ref{fig:ns}b, that achieves the same transformation on the space of zero, one and two photons, with the same probability of success. However, the two implementations do differ in the way they respond to variations in the transmission coefficients on the beam splitters and path length deviations. In this paper we analyse the difference in performance under these systematic errors of the two incarnations of the NS gate. We will refer to the original gate in Fig.~\ref{fig:ns}a as the KLM NS gate, and to Fig.~\ref{fig:ns}b as the Reverse NS gate.

The NS gate is \emph{nonlinear} in the sense that no combination of linear optical elements can implement the transformation in Eq.~(\ref{eq:ns}). The gate is induced by interference with ancilla photons in additional modes, followed by post-selection on a particular measurement outcome in the ancilla modes. This implies that the NS gate is successfully implemented with a probability smaller than one. The traditional implementation employs a single ancilla photon and two extra optical modes \cite{klm01}. The maximum success probability of any NS gate is $\smash{p_{\rm success} = \frac14}$ \cite{scheel04,eisert05}, which is achieved by both implementations in Fig.~\ref{fig:ns}. 

Next, we establish our conventions in describing the NS gates. We define the action of a beam splitter as a matrix transformation $U_{\rm BS}$ on the mode operators $\hat{a}_1$ and $\hat{a}_2$ of the two input modes $a_1$ and $a_2$, such that 
\begin{align}\label{eq:bs}
 U_{\rm BS}\, \hat{a}_1\, U_{\rm BS}^\dagger & = \cos\theta\,  \hat{a}_1 + \sin\theta\,  \hat{a}_2\, , \cr 
 U_{\rm BS}\, \hat{a}_2\, U_{\rm BS}^\dagger & =  - \sin\theta\,  \hat{a}_1 + \cos\theta\,  \hat{a}_2\, ,
\end{align}
and the mode operators are defined by the usual commutation relations 
\begin{align}
 \left[ \hat{a}_j , \hat{a}_k^\dagger \right] = \delta_{jk}\, ,
\end{align}
with $\delta_{jk}$ the Kronecker symbol. All other commutators are zero. The KLM NS gate in Fig.~\ref{fig:ns}a has beam splitter angles 
\begin{align}
 \theta_1 = \arccos\eta_1\, , \quad \theta_2 = \arccos\eta_2\, , \quad \theta_3 = -\theta_1\, , 
\end{align}
where 
\begin{align}
 \eta_1 = \frac{1}{4-2\sqrt{2}} \quad\text{and}\quad
 \eta_2 = 3-2\sqrt{2}\, .
\end{align}
The phases $\varphi_j$ in Fig.~\ref{fig:ns}a are all zero \cite{klm01}. We determine the beam splitter angles for the Reverse NS gate by requiring that the success probability of the ideal gate is again one quarter, and that the ancilla state and detection signature is the same as the KLM NS gate. Since we are primarily interested in finding an alternative gate and at this point do not wish to generate a complete family of NS gates, this construction suffices. We construct the mode transformations from Eq.~(\ref{eq:bs}) and collect the terms that have a single creation operator $\smash{\hat{a}_2^\dagger}$ and no mode operators $\smash{\hat{a}_3^\dagger}$, corresponding to the post-selection on a detected photon in mode $a_2$ and no detected photons in mode $a_3$. We then obtain coefficients $c_0$, $c_1$, and $c_2$ for the zero, one, and two-photon terms in the output state, respectively. Solving for $c_0 = c_1 = -c_2$, we obtain
\begin{align}
 \xi_1 = \arctan{\chi_1} \, , \quad \xi_2 =\pi + \arctan{\chi_2}\, , \quad \xi_3 = -\xi_1\, , 
\end{align}
where 
\begin{align}
 \chi_1 = \sqrt[4]{8} \qquad\text{and}\qquad
 \chi_2 = \frac{\sqrt{16\sqrt{2}-13}}{7} \, ,
\end{align}
with all phases $\zeta_j$ in Fig.~\ref{fig:ns}b equal to zero. Post-selection is implemented by projecting the three-mode output state onto the state ${\ket{1,0}}_{23}$, which has exactly one photon in mode 2, and zero photons in mode 3. For these values, the coefficients $\abs{c_k} = \frac12$, yielding an overall success probability of a quarter, independent of the input state in mode $a_1$.

In the next section we consider imperfections in the beam splitter transmission coefficients and the path lengths in the interferometer. We do not consider the effect of imperfect photodetection and ancilla preparation here, since this has already been considered in detail before by Lund \emph{et al.}\ \cite{lund03}.

\section{Imperfect components}\label{sec:components}\noindent
In any practical implementation of the NS gate, there will be variations in the components, such as beam splitter reflectivities and path length differences that introduce unwanted phases in the optical modes. These are systematic errors that must be overcome by calibration of some sort, rather than quantum error correction codes. In this section we will determine the sensitivity of each of the NS gate to variations in the three beam splitters and five path lengths. To this end we use the gate fidelity as a figure of merit \cite{bowdery02,nielsen02}. We find that given a specified target gate fidelity, the tolerances of the optical components vary significantly.

\begin{figure}[t!]
\centering
 \includegraphics[width=6cm]{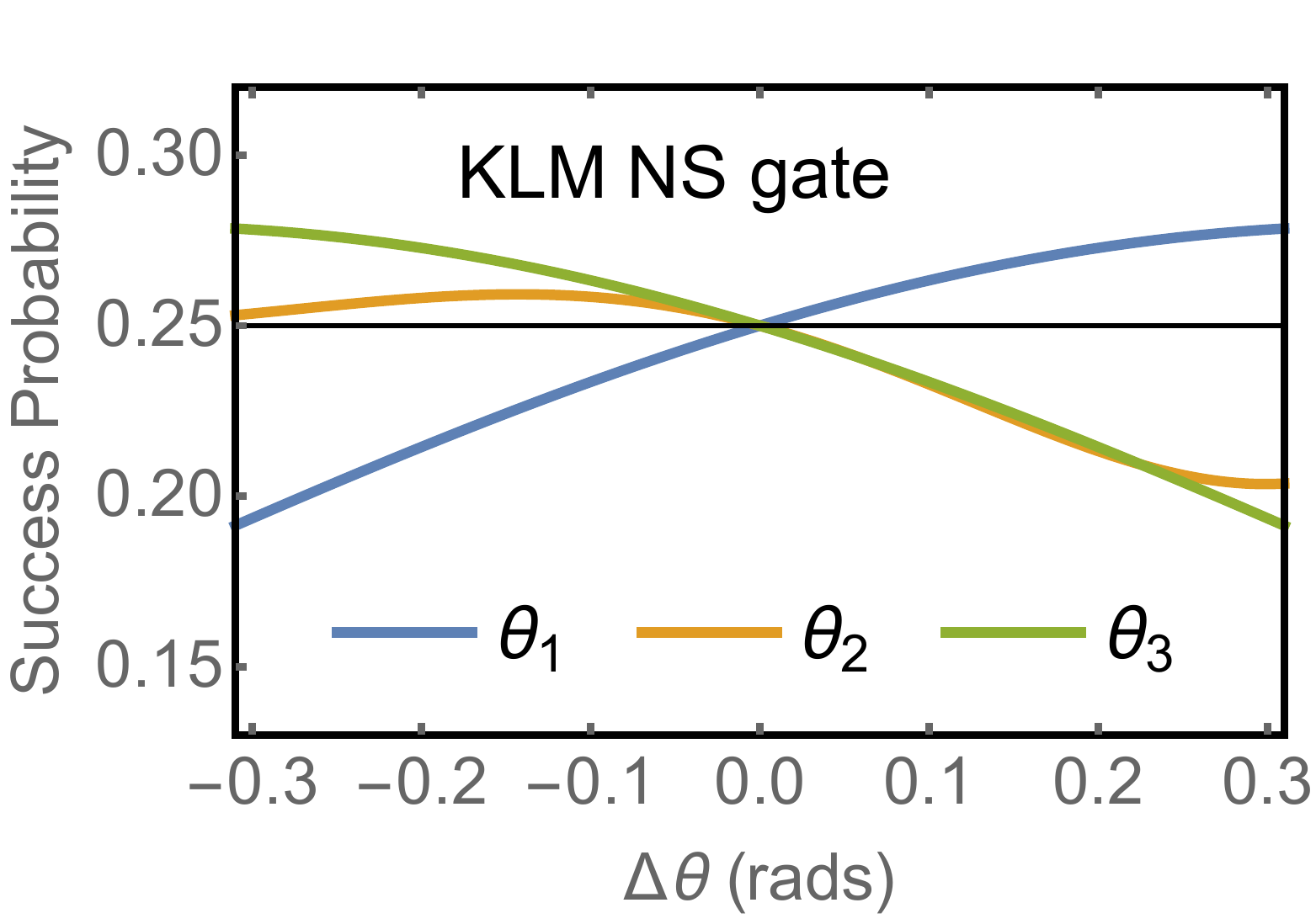} \\
 \includegraphics[width=6cm]{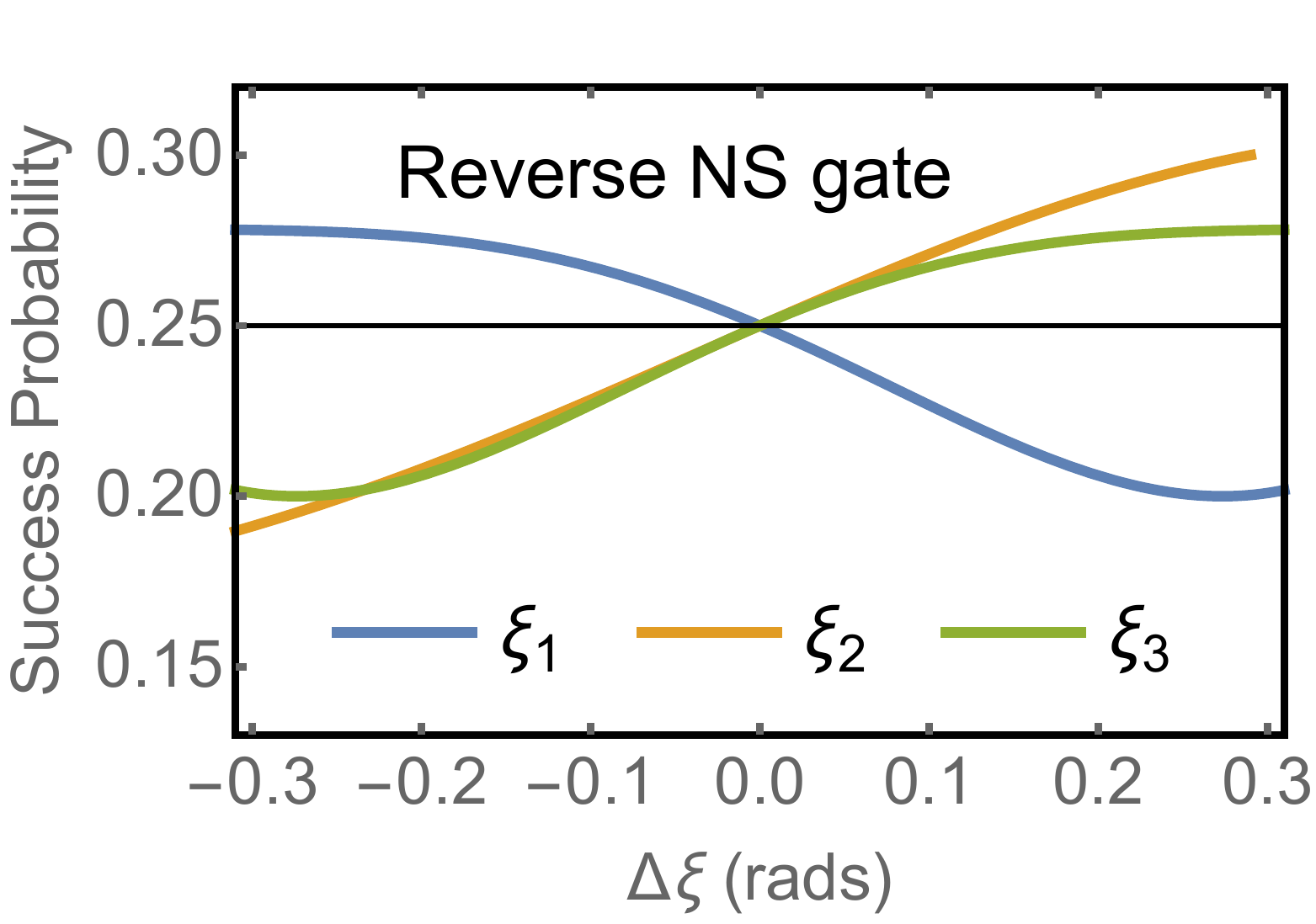}
 \caption{(Color online) The probability of a detector signature that heralds the success of the NS gates for varying beam splitter transmission coefficients and equal amplitudes $\alpha = \beta = \gamma = 1/\sqrt{3}$. In the top graph, $\Delta\theta_j$ is the variation away from the ideal value $\theta_j$ in the KLM NS gate, while in the bottom graph $\Delta\xi_j$ is the variation away from the ideal value $\xi_j$ in the Reverse NS gate. The black line indicates the ideal success probability of one quarter. Success probabilities larger that 0.25 are allowed, and indicate a significant departure from the ideal NS gate.}
\label{fig:nsprobs}
\end{figure}

Let $\mathcal{E}(\rho)$ denote a trace preserving quantum process on a density operator $\rho$. We define the gate fidelity of $\mathcal{E}$ relative to an ideal (unitary) gate $U$ as the quantity
\begin{align}\label{eq:gf}
 {F}(\mathcal{E},U) = \int d\psi\; \braket{\psi|U^\dagger\, \mathcal{E}(\psi)\, U|\psi}\, ,
\end{align}
where $d\psi$ is the uniform (Haar) measure over the quantum state space \cite{nielsen02}. Unfortunately, the NS gate is a non-trace-preserving quantum process, since the probability of success for the gate is less than one. Moreover, the success probability of the gate changes significantly with variations in the optical components. The probabilities of finding the detector signature that heralds success for the two NS gates as a function of variations in the beam splitter angles are shown in Fig.~\ref{fig:nsprobs}. While the theoretical maximum success probability of the ideal NS gate is one quarter, larger probabilities of finding the right detector outcomes are possible when the beam splitter coefficients change and the implemented gate deviates significantly from the ideal NS gate. We note that the curves for the first and third beam splitters are mirror images of each other in both the KLM and Reverse NS gate. This is explained by the time-reversal symmetric nature of the gates, keeping in mind that time-reversed detectors are sources, and vice versa. 

The variation in success probability means that we cannot use Eq.~(\ref{eq:gf}) in a straightforward manner. The process $\mathcal{E}$ must be normalised, but this means that Eq.~(\ref{eq:gf}) can no longer be evaluated analytically for the NS gates. Instead, we average the gate fidelity over 10\,000 random uniformly sampled input states for each value of gate component variations. Since our input state consists of a linear superposition of the first three Fock states (and ignoring a global phase), the state space is given by the unit sphere in a three-dimensional complex Hilbert space $\{ \psi \in \mathbb{C}^3 : ~ \norm{\psi} = 1\}$, where $\mathbb{C}$ is the complex plane and $\norm{\cdot}$ is the usual complex vector norm. Any state $\ket{\psi}$ in $\mathbb{C}^3$ can be obtained by applying a suitable matrix $U$ to an initial state $\ket{\psi_0}$, and finding a uniform distribution over the state space reduces to finding a uniform distribution over the set of unitary matrices $U$ acting on $\mathbb{C}^3$ with respect to the Haar measure. This is accomplished using the complex normal distribution on $\mathbb{C}^3$ \cite{nechita07}.

In the remainder of this section, we study the effect of beam splitter variations and path length differences individually, and calculate the minimum, maximum and mean gate fidelity for a distribution of variations across the circuit. 

\subsection{Imperfect beam splitters}
\begin{figure}[t!]
\centering
 \includegraphics[height=4.5cm]{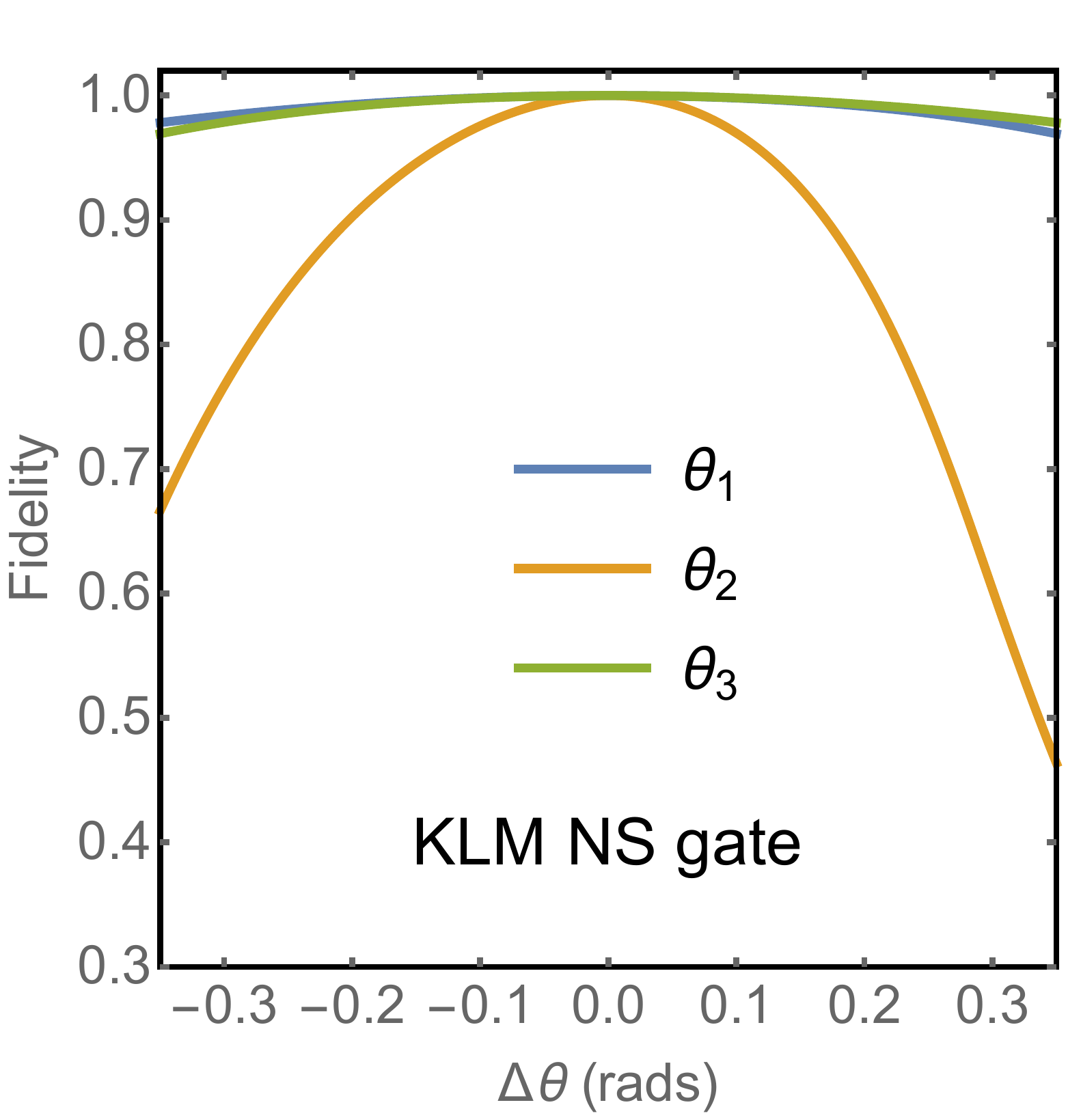}\!\!\!\!\!\!
 \includegraphics[height=4.5cm]{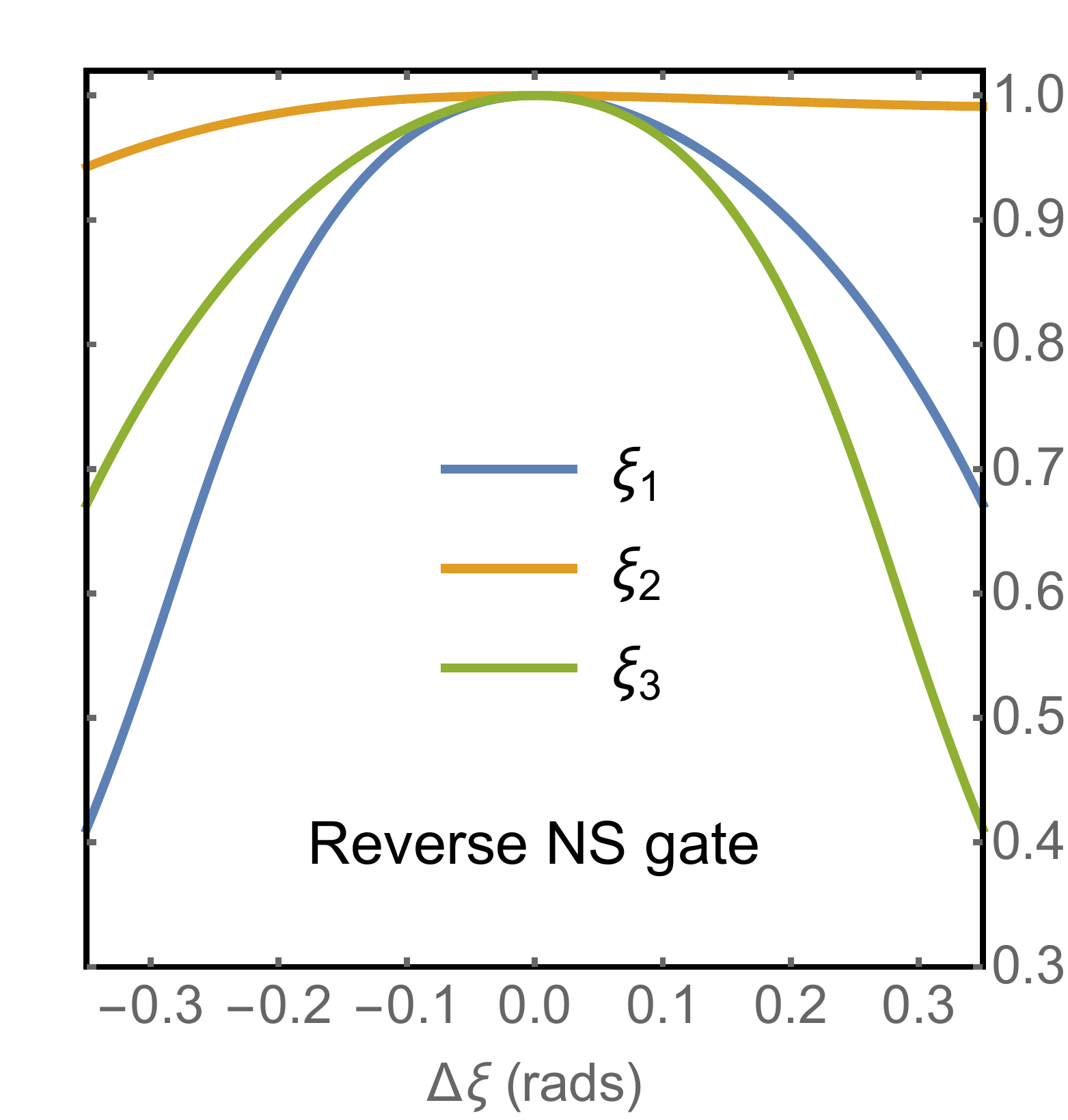}
\caption{(Color online) The gate fidelity of the KLM and Reverse NS gate for variations $\Delta\theta$ and $\Delta\xi$ in the three beam splitters, respectively. The Reverse NS gate is most sensitive to the first and third beam splitter, while the KLM NS gate is sensitive only to the second.}
\label{fig:bsvariations}
\end{figure}

\noindent
The gate fidelities for the KLM and Reverse NS gate with imperfect beam splitters are shown in Fig.~\ref{fig:bsvariations}. The KLM NS gate is particularly sensitive to variations in the second beam splitter, which is the one that directly interacts with the signal mode. The gate is significantly less sensitive to variations in the two other beam splitters. For example, at a fixed gate fidelity of $F=0.999$, the tolerance in the first and third beam splitters, ($\Delta\theta_1$ and $\Delta\theta_3$, respectively) is more than three times larger than $\Delta\theta_2$.  The curves for the first and third beam splitters are again mirror images of each other, as expected. The Reverse NS gate shows a similar range of sensitivities to beam splitter variations. Again, the two beam splitters in the direct signal path have the greatest effect on the gate fidelity, and the shape of these curves are very similar to the $\Delta\theta_2$ curve for the KLM NS gate. Again, we have mirror symmetry for the first and third beam splitters. 

Both gates have a success probability of $\smash{p_{\rm success} = \frac14}$, but the sensitivity of the Reverse NS gate of variations in two beam splitters instead of one makes the KLM NS gate the preferable circuit for practical implementations. In implementations that require high precision, the circuit design may require a variable beam splitter such as a directional coupler that is tuneable either mechanically \cite{Li09}, electrically \cite{Xu14}, or thermally \cite{Matthews09}. Since these structures are likely costly, and will introduce other imperfections (such as drift), the fewer beam splitters that need to be corrected in this way, the better. Generally, there will be room to optimise the circuit design based on the tolerances of the circuit on the variations in its elements. In practice, variations in beam splitters can be quite large in bulk optics (on the order of 10\%).

\subsection{Variations in path lengths}
\begin{figure}[t!]
\centering
 \includegraphics[height=4.5cm]{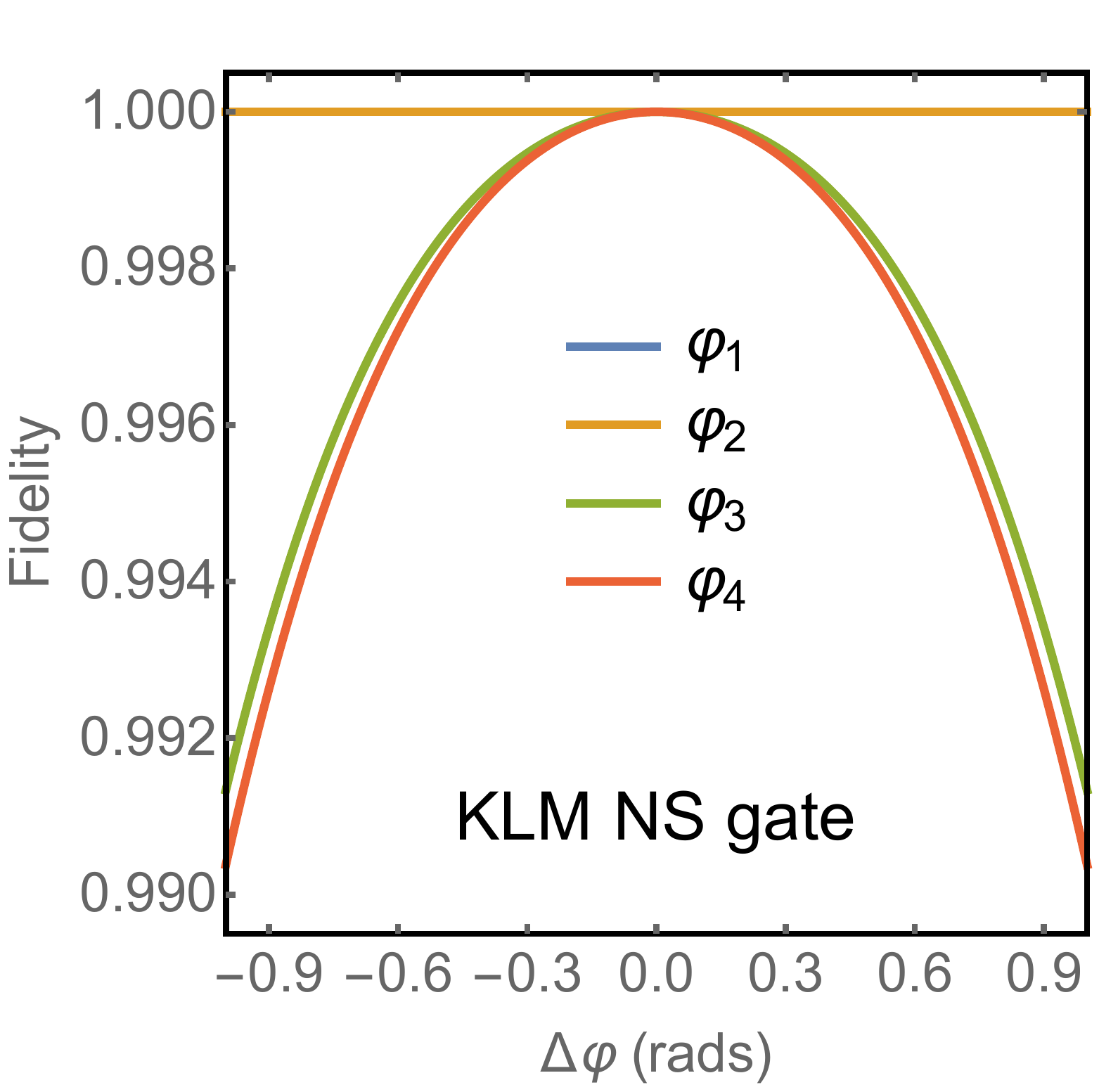}\!\!\!\!\!\!
 \includegraphics[height=4.5cm]{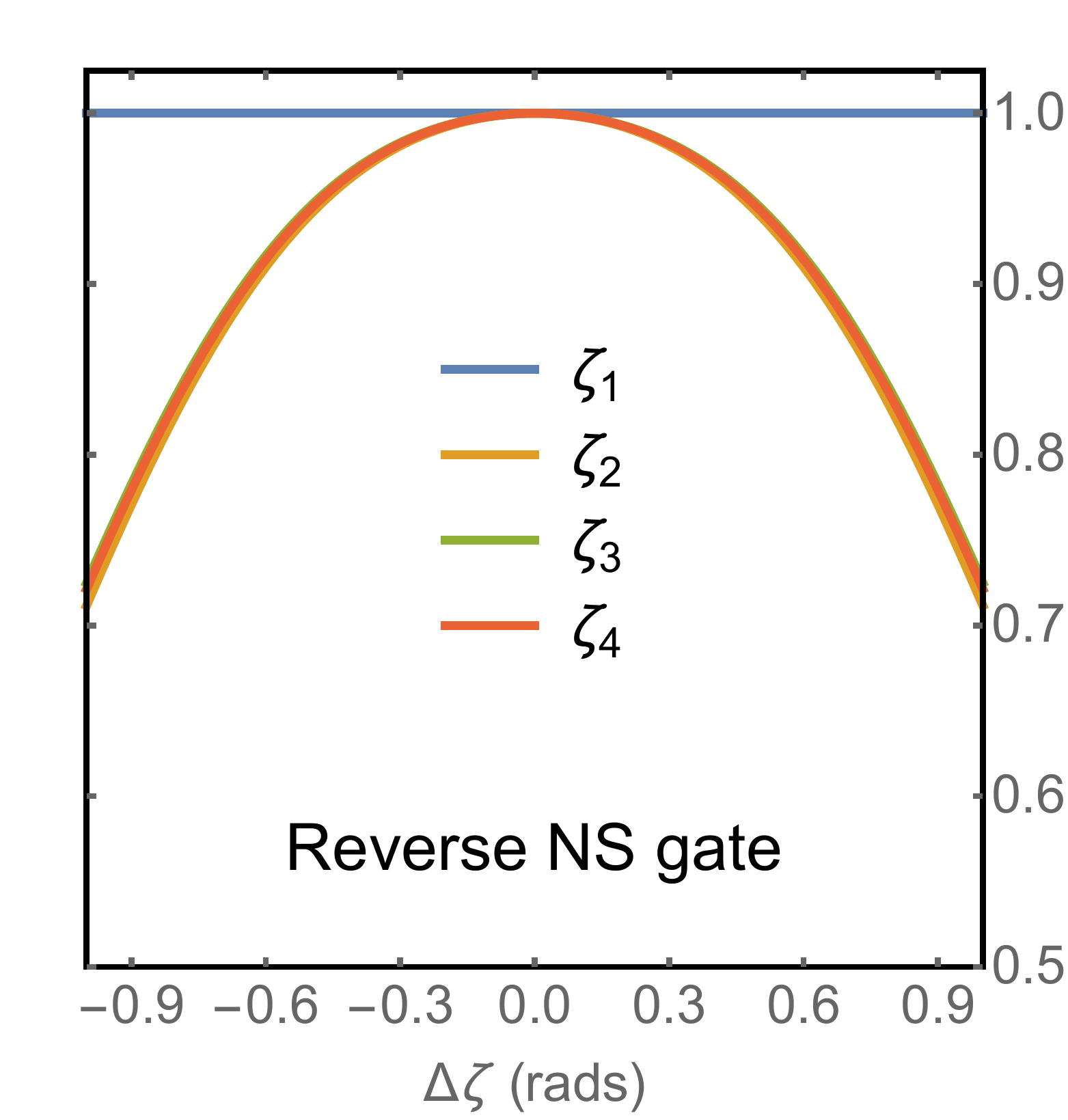}
\caption{(Color online) The gate fidelity of the KLM and Reverse NS gate for variations $\Delta\varphi_j$ and $\Delta\zeta_k$ in the five relevant path lengths, respectively. The KLM NS gate is again far less sensitive to path length variations than the Reverse NS gate.}
\label{fig:phase}
\end{figure}

\noindent
A similar analysis can be performed for the various path length differences in the circuit. The KLM NS gate is completely insensitive to path length variations encoded in the phases $\varphi_1$ and $\varphi_2$, as expected. Similarly, the Reverse NS gate is insensitive to variations in $\zeta_1$ (see Fig.~\ref{fig:phase}). For the remaining phases there is a marked difference in the two gates. The KLM NS gate loses only about a percent in fidelity when the path lengths associated with  $\varphi_3$ and  $\varphi_4$ varies by half a wavelength, while the Reverse gate sees a significant drop in average gate fidelity. Compared to the beam splitter variations we can say that the KLM NS gate is effectively insensitive to path length differences, while the Reverse NS gate is very sensitive to path length differences. This is another reason to strongly prefer the KLM NS gate over the Reverse NS gate, and underlines the importance of circuit analysis for component variations.

\subsection{Compound variations}
\begin{figure}[t!]
\centering
 \includegraphics[width=8.5cm]{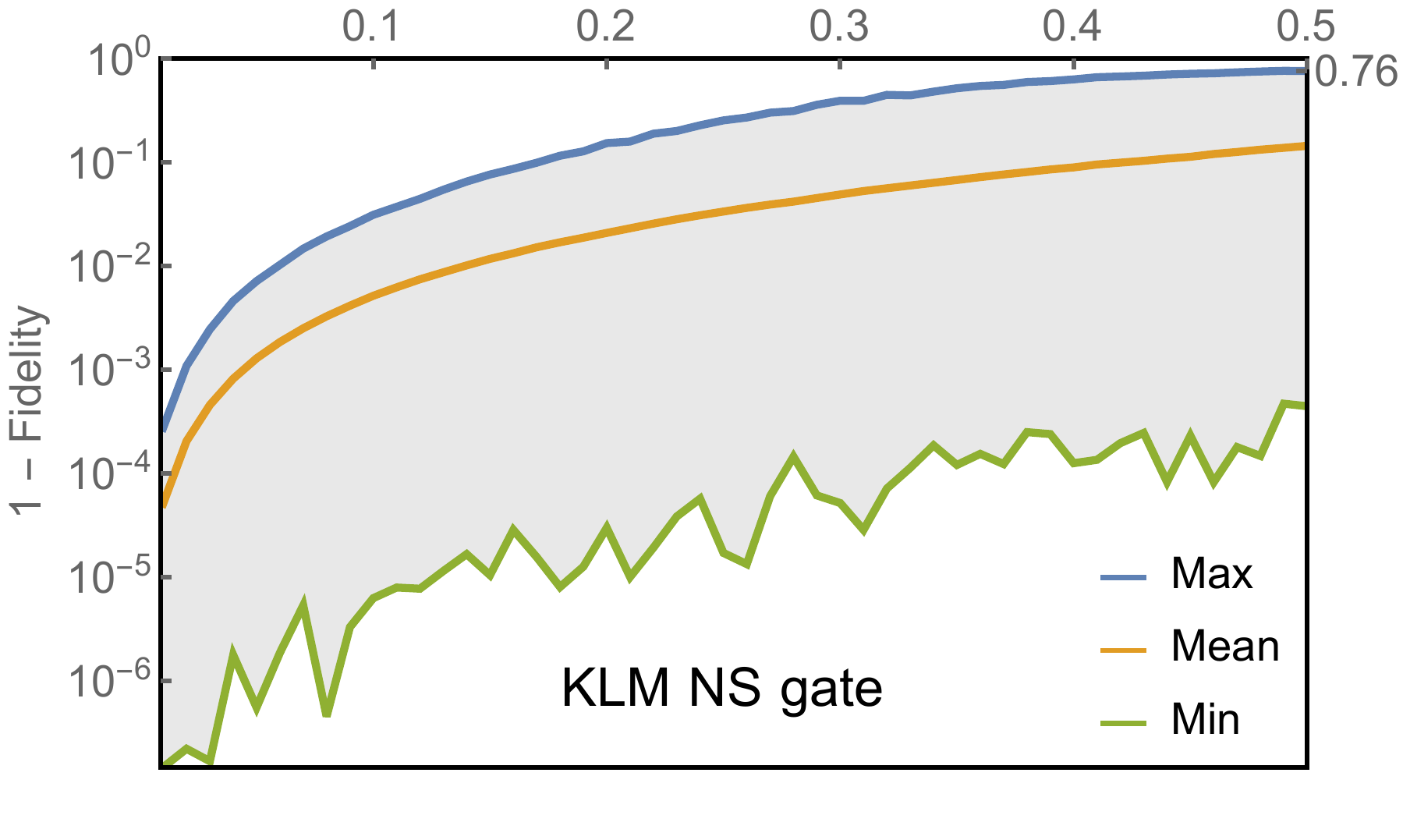} \\ \vskip -3mm
 \includegraphics[width=8.5cm]{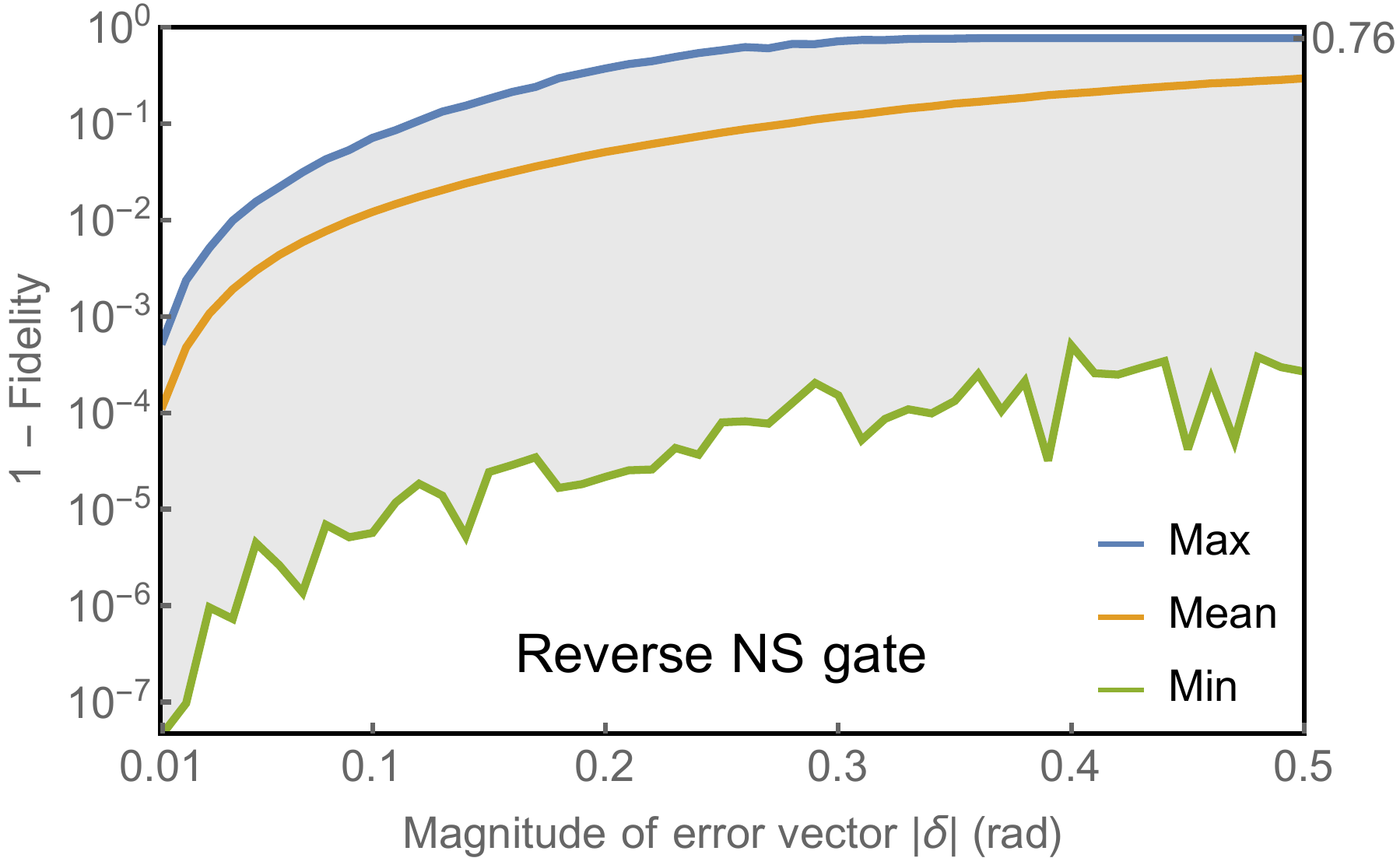}
\caption{(Color online) Minimum (green), maximum (blue), and mean (orange) gate infidelity as a function of the compound error $\abs{\bm{\delta}}$ for the KLM (top) and Reverse (bottom) NS gates. The minimum gate fidelity exhibits strong statistical fluctuations due to the relatively long tail in the distribution of gate fidelities for a given $\abs{\bm{\delta}}$. Both circuits reach the same randomisation limit of 0.76.}
\label{fig:compoundall}
\end{figure}

\noindent
In addition to individual errors in components, we may consider compound errors in all beam splitters and path lengths. This describes the more realistic behaviour where all components are subject to small variations. We define an error vector $\bm{\delta}$ with eight components (three for the beam splitters and five for the path lengths) and magnitude $\abs{\bm{\delta}}$. For a given total error $r$ distributed among the components, all possible error configurations are described by a 7-sphere of radius $\abs{\bm{\delta}}$. We randomly sample this error space and calculate the gate fidelity. The results are shown in Fig.~\ref{fig:compoundall}.

For each value of $\abs{\bm{\delta}}$ we generate 50\,000 random vectors $\bm{\delta}$. The gate fidelity for each $\bm{\delta}$ is then calculated again using 10\,000 random input states into the NS gate. In Fig.~\ref{fig:compoundall} we plot the minimum, maximum and mean gate infidelity $1-F$ as a function of $\abs{\bm{\delta}}$. The maximum gate infidelity is the worst case scenario for a given $\abs{\bm{\delta}}$. It reaches a maximum of approximately 0.76 with increasing $\abs{\bm{\delta}}$ for both the KLM and Reverse NS gate. This value represents the limit where the circuit is so badly constructed that it no longer outperforms a randomly constructed circuit, and we call this the randomisation limit. The gate infidelity of the Reverse NS gate reaches this value much faster than the KLM NS gate, which is consistent with our earlier observation that the Reverse NS gate is more sensitive to variations in the two in-line beam splitters, compared to the KLM NS gate that is sensitive to variations in the single in-line beam splitter. In both cases the effect of path length variations is much less important than the beam splitter variations.

The minimum gate infidelity exhibits strong statistical fluctuations (the green lines in Fig.~\ref{fig:compoundall}). This is due to the long tail in the fidelity distribution for given $\abs{\bm{\delta}}$. Only relatively few circuit configurations $\bm{\delta}$ will give a high gate fidelity, and the finite number of samples (50\,000) is unlikely to hit upon the true minimum gate infidelity. This line is therefore more accurately characterised as a lower bound on the maximum gate fidelity (or, equivalently, an upper bound on the minimum gate infidelity).

The method used to analyse compound errors can be generalised for any system, but in practice this is a computationally intensive process. Generating $k$ error vectors over $n$ input states requires $kn$ evaluations of the gate fidelity. For 3-port optical networks this is still tractable, but for general $N$-port networks the number of optical elements---and therefore the dimension of the error vector---scales as $\mathcal{O}(N^2)$. We will need more sophisticated theoretical methods to analyse complex optical networks.

\section{Quantum estimation of Circuit Components}\label{sec:multi}\noindent
One potentially fruitful approach towards analysing the effect of component variations on the gate fidelity in linear optical networks is to use the theory of multi-parameter quantum estimation. When we consider the components of the multi-mode interferometer, the variations away from the ideal designed value become a vector of random variables $\bm{\delta}$ in a parameter estimation problem. We can then use techniques from quantum metrology \cite{giovannetti11}, information geometry \cite{amari93}, and the theory of the dynamical evolution of quantum states \cite{jones10} to shed light on the sensitivity of an interferometer on its elements. 

\begin{figure}[t!]
\centering
 \includegraphics[width=6cm]{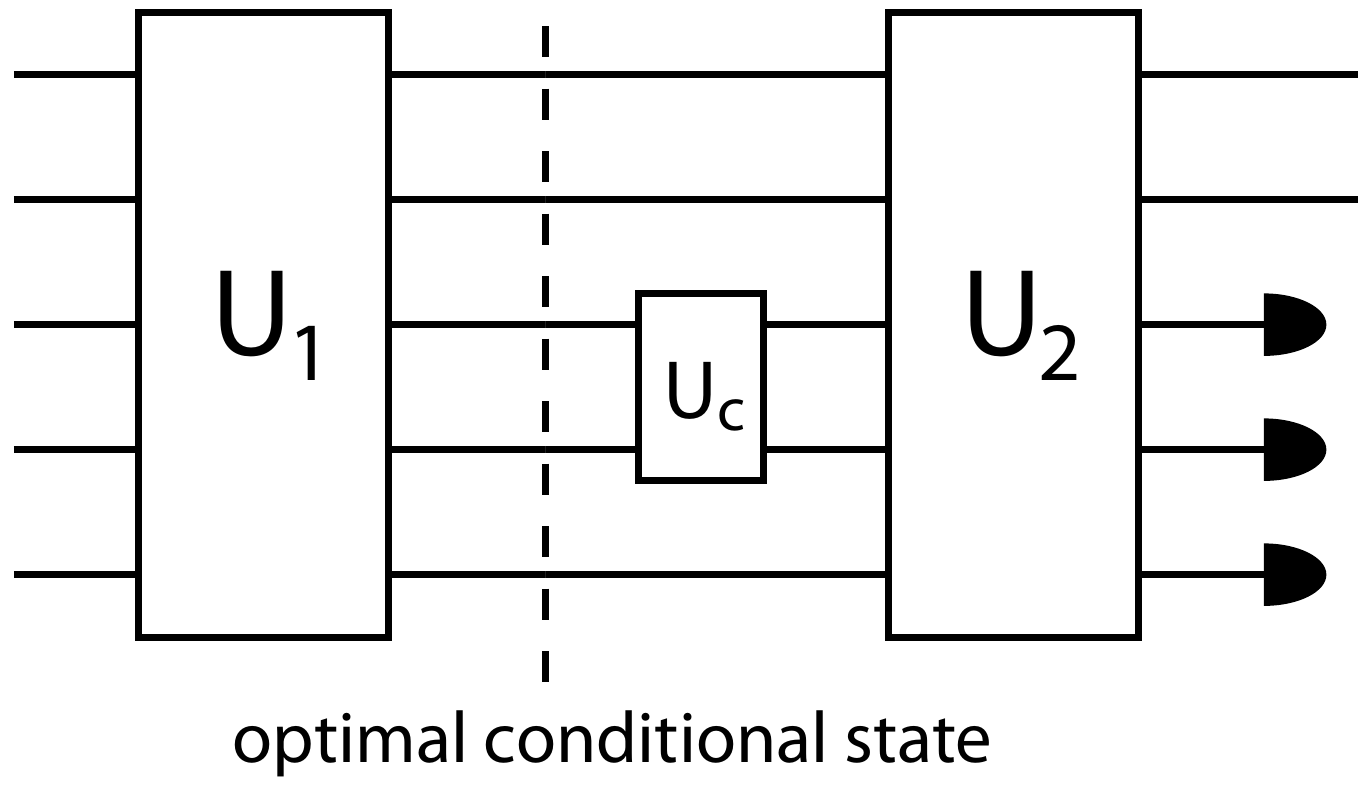}
\caption{The optimal conditional state $\ket{\Psi_c^{\rm opt}}$ is taken just before the component $c$ of interest (dashed line), and it is the state that maximises the quantum Fisher information (evaluated by the variance of the generator of $U_c$ with respect to $\ket{\Psi_c^{\rm opt}}$), while still being capable of triggering the detector array in the required way.}
\label{fig:qfi}
\end{figure}

The quantum Fisher information (QFI) is a metric in the state space that is parametrised by the random variables $\bm{\delta}$. It is a special case of the Bures metric \cite{helstrom67}.  Intuitively, the quantum Fisher information $I_Q(\bm{\delta})$ is the amount of information about $\bm{\delta}$ that is contained in the state $\ket{\psi}$. However, for our purposes it is sufficient to note that a large QFI means that we can detect small variations in $\bm{\delta}$. Therefore, $I_Q(\bm{\delta})$ is also a metric for the sensitivity of $\ket{\psi}$ on $\bm{\delta}$.

Let a unitary transformation of an optical circuit be denoted by $U$, and deviations in the characteristics of a component $c$ (such as a beam splitter or phase shifter) are generated by $G_c$. The unitary transformation corresponding to an inaccurate component is then  
\begin{align}
 \widetilde{U}_{c}(\delta_c) = \exp(-iG_c \delta_c)\, U_c \exp(iG_c \delta_c) \, ,
\end{align}
where $\delta_c$ is a component of the vector $\bm{\delta}$. The QFI for $\delta_c$ is bounded by the variance $(\Delta G_c)^2$ of $G_c$ with respect to the optical quantum state $\ket{\Psi_c}$ immediately prior to the component $c$ \cite{braunstein95}. By studying $(\Delta G_c)^2$ with respect to a variety of quantum states $\ket{\Psi_c}$ (average, best and worst case scenario) we can estimate the effect of variations of that component on the total gate. Post-selection on a particular detection signature (as in the case of the NS gate) will typically exclude certain states $\ket{\Psi_c}$, and the most informative average QFI will no longer be due to a uniform distribution of $\ket{\Psi_c}$ in the quantum state space. Instead, the set of $\ket{\Psi_c}$ that are to be averaged over should be constructed from a uniform distribution of input states over all the non-ancilla input modes, tensored with the ancilla input states and transformed to the state just before the component $c$.

The above procedure still requires averaging over a large number of states. To circumvent this lengthy process, we need a way to determine the optimal conditional state that maximises the QFI as evaluated by $(\Delta G_c)^2$, while still being capable of triggering the detectors according to the required signature (see Fig.~\ref{fig:qfi}). The variance $(\Delta G_c)^2$ must be evaluated with respect to this optimal state $\ket{\Psi_c^{\rm opt}}$. Again, this may be a computationally difficult problem.

Finally, we can calculate the weighted average $W_c$ over the variance $(\Delta G_c)^2_\Psi$ as a measure of the sensitivity of a component to variations:
\begin{align}
 W_c = \int d\Psi\; p_\Psi (\Delta G_c)^2_\Psi  \, ,
\end{align}
where $p_\Psi$ is the probability that the state $\ket{\Psi_c}$ leads to the required detector signature, and $d\Psi$ is the Haar measure over the entire multi-mode optical state just before entering component $c$. For the NS gate this is a three-mode state. We also explicitly included the subscript on $(\Delta G_c)^2_\Psi$ to remind ourselves that the variance depends on the input state. Which of these approaches is most suitable likely depends on the specifics of the optical circuit under consideration, and the exact relation between circuit analysis and quantum parameter estimation will be the subject of future studies.

\section{Discussion and Conclusions}\label{sec:disc}\noindent
We have shown that the construction of optical networks that implement a given unitary transformation is generally not unique, and that variations in the components of the network can have dramatically different effects on the network (gate) fidelity. Moreover, different network topologies for the same transformation may place very different precision requirements on the components, and any practical implementation should involve a circuit analysis on how to best implement the optical network. For small networks, a simple numerical calculation of the average gate fidelity may be tractable, but larger networks require more sophisticated methods. One such method is the quantum Fisher information, which can be calculated efficiently for optical components by considering the variance of the generator of translations.

Our findings prompt a number of important questions for future research into the practical construction of optical networks:
 (i) Why do some elements in a network require much more precise fabrication than others? 
 (ii) How can we design optical networks that minimise the number of sensitive elements? 
 (iii) How can we determine the component characteristics \emph{in situ}, after the network has been fabricated? 
These questions will be studied further in future work.

\section*{Acknowledgements}\noindent
PK thanks Mercedes Gimeno-Segovia for stimulating discussions on alternative implementations of the NS gate.


\begin{thebibliography}{18}%
\makeatletter
\providecommand \@ifxundefined [1]{%
 \@ifx{#1\undefined}
}%
\providecommand \@ifnum [1]{%
 \ifnum #1\expandafter \@firstoftwo
 \else \expandafter \@secondoftwo
 \fi
}%
\providecommand \@ifx [1]{%
 \ifx #1\expandafter \@firstoftwo
 \else \expandafter \@secondoftwo
 \fi
}%
\providecommand \natexlab [1]{#1}%
\providecommand \enquote  [1]{``#1''}%
\providecommand \bibnamefont  [1]{#1}%
\providecommand \bibfnamefont [1]{#1}%
\providecommand \citenamefont [1]{#1}%
\providecommand \href@noop [0]{\@secondoftwo}%
\providecommand \href [0]{\begingroup \@sanitize@url \@href}%
\providecommand \@href[1]{\@@startlink{#1}\@@href}%
\providecommand \@@href[1]{\endgroup#1\@@endlink}%
\providecommand \@sanitize@url [0]{\catcode `\\12\catcode `\$12\catcode
  `\&12\catcode `\#12\catcode `\^12\catcode `\_12\catcode `\%12\relax}%
\providecommand \@@startlink[1]{}%
\providecommand \@@endlink[0]{}%
\providecommand \url  [0]{\begingroup\@sanitize@url \@url }%
\providecommand \@url [1]{\endgroup\@href {#1}{\urlprefix }}%
\providecommand \urlprefix  [0]{URL }%
\providecommand \Eprint [0]{\href }%
\providecommand \doibase [0]{http://dx.doi.org/}%
\providecommand \selectlanguage [0]{\@gobble}%
\providecommand \bibinfo  [0]{\@secondoftwo}%
\providecommand \bibfield  [0]{\@secondoftwo}%
\providecommand \translation [1]{[#1]}%
\providecommand \BibitemOpen [0]{}%
\providecommand \bibitemStop [0]{}%
\providecommand \bibitemNoStop [0]{.\EOS\space}%
\providecommand \EOS [0]{\spacefactor3000\relax}%
\providecommand \BibitemShut  [1]{\csname bibitem#1\endcsname}%
\let\auto@bib@innerbib\@empty
\bibitem [{\citenamefont {Caulfield}\ and\ \citenamefont
  {Dolev}(2010)}]{shlomi10}%
  \BibitemOpen
  \bibfield  {author} {\bibinfo {author} {\bibfnamefont {H.~J.}\ \bibnamefont
  {Caulfield}}\ and\ \bibinfo {author} {\bibfnamefont {S.}~\bibnamefont
  {Dolev}},\ }\href {\doibase 10.1038/nphoton.2010.94} {\bibfield  {journal}
  {\bibinfo  {journal} {Nature Photon.}\ }\textbf {\bibinfo {volume} {4}},\
  \bibinfo {pages} {261} (\bibinfo {year} {2010})}\BibitemShut {NoStop}%
\bibitem [{\citenamefont {Knill}\ \emph {et~al.}(2001)\citenamefont {Knill},
  \citenamefont {Laflamme},\ and\ \citenamefont {Milburn}}]{klm01}%
  \BibitemOpen
  \bibfield  {author} {\bibinfo {author} {\bibfnamefont {E.}~\bibnamefont
  {Knill}}, \bibinfo {author} {\bibfnamefont {R.}~\bibnamefont {Laflamme}}, \
  and\ \bibinfo {author} {\bibfnamefont {G.~J.}\ \bibnamefont {Milburn}},\
  }\href {\doibase 10.1038/35051009} {\bibfield  {journal} {\bibinfo  {journal}
  {Nature}\ }\textbf {\bibinfo {volume} {409}},\ \bibinfo {pages} {46}
  (\bibinfo {year} {2001})}\BibitemShut {NoStop}%
\bibitem [{\citenamefont {Kok}\ \emph {et~al.}(2007)\citenamefont {Kok},
  \citenamefont {Nemoto}, \citenamefont {Ralph}, \citenamefont {Dowling},\ and\
  \citenamefont {Milburn}}]{kok07}%
  \BibitemOpen
  \bibfield  {author} {\bibinfo {author} {\bibfnamefont {P.}~\bibnamefont
  {Kok}}, \bibinfo {author} {\bibfnamefont {K.}~\bibnamefont {Nemoto}},
  \bibinfo {author} {\bibfnamefont {T.~C.}\ \bibnamefont {Ralph}}, \bibinfo
  {author} {\bibfnamefont {J.~P.}\ \bibnamefont {Dowling}}, \ and\ \bibinfo
  {author} {\bibfnamefont {G.~J.}\ \bibnamefont {Milburn}},\ }\href {\doibase
  10.1103/RevModPhys.79.135} {\bibfield  {journal} {\bibinfo  {journal} {Rev.
  Mod. Phys.}\ }\textbf {\bibinfo {volume} {79}},\ \bibinfo {pages} {135}
  (\bibinfo {year} {2007})}\BibitemShut {NoStop}%
\bibitem [{\citenamefont {Kok}\ and\ \citenamefont
  {Lovett}(2010)}]{koklovett10}%
  \BibitemOpen
  \bibfield  {author} {\bibinfo {author} {\bibfnamefont {P.}~\bibnamefont
  {Kok}}\ and\ \bibinfo {author} {\bibfnamefont {B.~W.}\ \bibnamefont
  {Lovett}},\ }\href {\doibase 10.1017/CBO9781139193658} {\emph {\bibinfo
  {title} {Introduction to Optical Quantum Information Processing}}}\ (\bibinfo
   {publisher} {Cambridge},\ \bibinfo {year} {2010})\BibitemShut {NoStop}%
\bibitem [{\citenamefont {Scheel}\ and\ \citenamefont
  {Lutkenhaus}(2004)}]{scheel04}%
  \BibitemOpen
  \bibfield  {author} {\bibinfo {author} {\bibfnamefont {S.}~\bibnamefont
  {Scheel}}\ and\ \bibinfo {author} {\bibfnamefont {N.}~\bibnamefont
  {Lutkenhaus}},\ }\href {\doibase 10.1088/1367-2630/6/1/051} {\bibfield
  {journal} {\bibinfo  {journal} {New J. Phys.}\ }\textbf {\bibinfo {volume}
  {6}},\ \bibinfo {pages} {51} (\bibinfo {year} {2004})}\BibitemShut {NoStop}%
\bibitem [{\citenamefont {Eisert}(2005)}]{eisert05}%
  \BibitemOpen
  \bibfield  {author} {\bibinfo {author} {\bibfnamefont {J.}~\bibnamefont
  {Eisert}},\ }\href {\doibase http://dx.doi.org/10.1103/PhysRevLett.95.040502}
  {\bibfield  {journal} {\bibinfo  {journal} {Phys. Rev. Lett.}\ }\textbf
  {\bibinfo {volume} {95}},\ \bibinfo {pages} {040502} (\bibinfo {year}
  {2005})}\BibitemShut {NoStop}%
\bibitem [{\citenamefont {Lund}\ \emph {et~al.}(2003)\citenamefont {Lund},
  \citenamefont {Bell},\ and\ \citenamefont {Ralph}}]{lund03}%
  \BibitemOpen
  \bibfield  {author} {\bibinfo {author} {\bibfnamefont {A.~P.}\ \bibnamefont
  {Lund}}, \bibinfo {author} {\bibfnamefont {T.~B.}\ \bibnamefont {Bell}}, \
  and\ \bibinfo {author} {\bibfnamefont {T.~C.}\ \bibnamefont {Ralph}},\ }\href
  {\doibase 10.1103/PhysRevA.68.022313} {\bibfield  {journal} {\bibinfo
  {journal} {Phys. Rev. A}\ }\textbf {\bibinfo {volume} {68}},\ \bibinfo
  {pages} {022313} (\bibinfo {year} {2003})}\BibitemShut {NoStop}%
\bibitem [{\citenamefont {Bowdrey}\ \emph {et~al.}(2002)\citenamefont
  {Bowdrey}, \citenamefont {Oi}, \citenamefont {Short}, \citenamefont
  {Banaszek},\ and\ \citenamefont {Jones}}]{bowdery02}%
  \BibitemOpen
  \bibfield  {author} {\bibinfo {author} {\bibfnamefont {M.~D.}\ \bibnamefont
  {Bowdrey}}, \bibinfo {author} {\bibfnamefont {D.}~\bibnamefont {Oi}},
  \bibinfo {author} {\bibfnamefont {A.~J.}\ \bibnamefont {Short}}, \bibinfo
  {author} {\bibfnamefont {K.}~\bibnamefont {Banaszek}}, \ and\ \bibinfo
  {author} {\bibfnamefont {J.~A.}\ \bibnamefont {Jones}},\ }\href {\doibase
  10.1016/S0375-9601(02)00069-5} {\bibfield  {journal} {\bibinfo  {journal}
  {Phys. Lett. A}\ }\textbf {\bibinfo {volume} {294}},\ \bibinfo {pages} {258}
  (\bibinfo {year} {2002})}\BibitemShut {NoStop}%
\bibitem [{\citenamefont {Nielsen}(2002)}]{nielsen02}%
  \BibitemOpen
  \bibfield  {author} {\bibinfo {author} {\bibfnamefont {M.~A.}\ \bibnamefont
  {Nielsen}},\ }\href {\doibase 10.1016/S0375-9601(02)01272-0} {\bibfield
  {journal} {\bibinfo  {journal} {Phys. Rev. A}\ }\textbf {\bibinfo {volume}
  {303}},\ \bibinfo {pages} {249} (\bibinfo {year} {2002})}\BibitemShut
  {NoStop}%
\bibitem [{\citenamefont {Nechita}(2007)}]{nechita07}%
  \BibitemOpen
  \bibfield  {author} {\bibinfo {author} {\bibfnamefont {I.}~\bibnamefont
  {Nechita}},\ }\href {\doibase 0.1007/s00023-007-0345-5} {\bibfield  {journal}
  {\bibinfo  {journal} {{Ann. H. Poincar\'e}}\ }\textbf {\bibinfo {volume}
  {8}},\ \bibinfo {pages} {1521} (\bibinfo {year} {2007})}\BibitemShut
  {NoStop}%
\bibitem [{\citenamefont {Li}\ \emph {et~al.}(2009)\citenamefont {Li},
  \citenamefont {Pernice},\ and\ \citenamefont {Tang}}]{Li09}%
  \BibitemOpen
  \bibfield  {author} {\bibinfo {author} {\bibfnamefont {M.}~\bibnamefont
  {Li}}, \bibinfo {author} {\bibfnamefont {W.~H.~P.}\ \bibnamefont {Pernice}},
  \ and\ \bibinfo {author} {\bibfnamefont {H.~X.}\ \bibnamefont {Tang}},\
  }\href {\doibase 10.1038/nphoton.2009.116} {\bibfield  {journal} {\bibinfo
  {journal} {Nature Photon.}\ }\textbf {\bibinfo {volume} {3}},\ \bibinfo
  {pages} {464} (\bibinfo {year} {2009})}\BibitemShut {NoStop}%
\bibitem [{\citenamefont {Xu}\ \emph {et~al.}(2014)\citenamefont {Xu},
  \citenamefont {Chen}, \citenamefont {Wood}, \citenamefont {Sun},\ and\
  \citenamefont {Reano}}]{Xu14}%
  \BibitemOpen
  \bibfield  {author} {\bibinfo {author} {\bibfnamefont {Q.}~\bibnamefont
  {Xu}}, \bibinfo {author} {\bibfnamefont {L.}~\bibnamefont {Chen}}, \bibinfo
  {author} {\bibfnamefont {M.~G.}\ \bibnamefont {Wood}}, \bibinfo {author}
  {\bibfnamefont {P.}~\bibnamefont {Sun}}, \ and\ \bibinfo {author}
  {\bibfnamefont {R.~M.}\ \bibnamefont {Reano}},\ }\href {\doibase
  10.1038/ncomms6337} {\bibfield  {journal} {\bibinfo  {journal} {Nature
  Commun.}\ }\textbf {\bibinfo {volume} {5}},\ \bibinfo {pages} {346} (\bibinfo
  {year} {2014})}\BibitemShut {NoStop}%
\bibitem [{\citenamefont {Matthews}\ \emph {et~al.}(2009)\citenamefont
  {Matthews}, \citenamefont {Politi}, \citenamefont {Stefanov},\ and\
  \citenamefont {O'Brien}}]{Matthews09}%
  \BibitemOpen
  \bibfield  {author} {\bibinfo {author} {\bibfnamefont {J.~C.~F.}\
  \bibnamefont {Matthews}}, \bibinfo {author} {\bibfnamefont {A.}~\bibnamefont
  {Politi}}, \bibinfo {author} {\bibfnamefont {A.}~\bibnamefont {Stefanov}}, \
  and\ \bibinfo {author} {\bibfnamefont {J.~L.}\ \bibnamefont {O'Brien}},\
  }\href {\doibase 10.1038/nphoton.2009.93} {\bibfield  {journal} {\bibinfo
  {journal} {Nature Photon.}\ }\textbf {\bibinfo {volume} {3}},\ \bibinfo
  {pages} {346} (\bibinfo {year} {2009})}\BibitemShut {NoStop}%
\bibitem [{\citenamefont {Giovannetti}\ \emph {et~al.}(2011)\citenamefont
  {Giovannetti}, \citenamefont {Lloyd},\ and\ \citenamefont
  {Maccone}}]{giovannetti11}%
  \BibitemOpen
  \bibfield  {author} {\bibinfo {author} {\bibfnamefont {V.}~\bibnamefont
  {Giovannetti}}, \bibinfo {author} {\bibfnamefont {S.}~\bibnamefont {Lloyd}},
  \ and\ \bibinfo {author} {\bibfnamefont {L.}~\bibnamefont {Maccone}},\ }\href
  {\doibase 10.1038/nphoton.2011.35} {\bibfield  {journal} {\bibinfo  {journal}
  {Nature Phot.}\ }\textbf {\bibinfo {volume} {5}},\ \bibinfo {pages} {222}
  (\bibinfo {year} {2011})}\BibitemShut {NoStop}%
\bibitem [{\citenamefont {Amari}\ and\ \citenamefont
  {Nagaoka}(1993)}]{amari93}%
  \BibitemOpen
  \bibfield  {author} {\bibinfo {author} {\bibfnamefont {S.}~\bibnamefont
  {Amari}}\ and\ \bibinfo {author} {\bibfnamefont {H.}~\bibnamefont
  {Nagaoka}},\ }\href {\doibase 10.1109/TIT.2009.2016067} {\emph {\bibinfo
  {title} {{Methods in Information Geometry}}}}\ (\bibinfo  {publisher}
  {Oxford},\ \bibinfo {year} {1993})\BibitemShut {NoStop}%
\bibitem [{\citenamefont {Jones}\ and\ \citenamefont {Kok}(2010)}]{jones10}%
  \BibitemOpen
  \bibfield  {author} {\bibinfo {author} {\bibfnamefont {P.~J.}\ \bibnamefont
  {Jones}}\ and\ \bibinfo {author} {\bibfnamefont {P.}~\bibnamefont {Kok}},\
  }\href {\doibase 10.1103/PhysRevA.82.022107} {\bibfield  {journal} {\bibinfo
  {journal} {Phys. Rev. A}\ }\textbf {\bibinfo {volume} {82}},\ \bibinfo
  {pages} {022107} (\bibinfo {year} {2010})}\BibitemShut {NoStop}%
\bibitem [{\citenamefont {Helstrom}(1967)}]{helstrom67}%
  \BibitemOpen
  \bibfield  {author} {\bibinfo {author} {\bibfnamefont {C.~W.}\ \bibnamefont
  {Helstrom}},\ }\href {\doibase 10.1016/0375-9601(67)90366-0} {\bibfield
  {journal} {\bibinfo  {journal} {Phys. Lett. A}\ }\textbf {\bibinfo {volume}
  {25}},\ \bibinfo {pages} {101} (\bibinfo {year} {1967})}\BibitemShut
  {NoStop}%
\bibitem [{\citenamefont {Braunstein}\ \emph {et~al.}(1995)\citenamefont
  {Braunstein}, \citenamefont {Caves},\ and\ \citenamefont
  {Milburn}}]{braunstein95}%
  \BibitemOpen
  \bibfield  {author} {\bibinfo {author} {\bibfnamefont {S.~L.}\ \bibnamefont
  {Braunstein}}, \bibinfo {author} {\bibfnamefont {C.~M.}\ \bibnamefont
  {Caves}}, \ and\ \bibinfo {author} {\bibfnamefont {G.~J.}\ \bibnamefont
  {Milburn}},\ }\href {\doibase 10.1006/aphy.1996.0040} {\bibfield  {journal}
  {\bibinfo  {journal} {Ann. Phys.}\ }\textbf {\bibinfo {volume} {247}},\
  \bibinfo {pages} {135} (\bibinfo {year} {1995})}\BibitemShut {NoStop}%
\end{thebibliography}
\end{document}